\newcommand{\lpx}{\stackrel{\leftarrow}{\partial}_{x}}
\newcommand{\lpy}{\stackrel{\leftarrow}{\partial}_{y}}
\newcommand{\rpx}{\stackrel{\rightarrow}{\partial}_{x}}
\newcommand{\rpy}{\stackrel{\rightarrow}{\partial}_{x}}
\newcommand{\lpp}{\stackrel{\leftarrow}{\partial}_{p}}
\newcommand{\rpp}{\stackrel{\rightarrow}{\partial}_{p}}

\newcommand{\ben}{\begin{equation}}
\newcommand{\een}{\end{equation}}
\newcommand{\bea}{\begin{eqnarray}}
\newcommand{\eea}{\end{eqnarray}}
\newcommand{\nn}{\nonumber\\ }

\newcommand{\re}{{\rm Re}}
\newcommand{\im}{{\rm Im}}

\renewcommand{\*}{\star}

\newcommand{\qq}{\qquad\qquad}
\newcommand{\QQ}{\qquad\qquad\qquad\qquad}
\newcommand{\erf}{{\rm erf}}
\newcommand{\erfc}{{\rm erfc}}
\newcommand{\infinity}{\infty}

\newcommand{\ssee}{$\*$-eigen-$\*$ equation}
\documentclass{article}

\begin{document}

\parskip=4pt
\baselineskip=14pt


\title{Wigner functions, contact interactions, and matching\\
}
\author{Mark A. Walton\\ \\ {\em Department of Physics,
University of Lethbridge}\\
{\em Lethbridge, Alberta, Canada\ \  T1K 3M4}\\
{\small walton@uleth.ca}\\ \\
}

\maketitle
\begin{abstract}
Quantum mechanics in phase space (or deformation quantization)
appears to fail as an autonomous quantum method when infinite
potential walls are present. The stationary physical Wigner
functions do not satisfy the normal eigen equations, the
$\*$-eigen equations, unless an ad hoc boundary potential is added
\cite{DP}. Alternatively, they satisfy a different, higher-order,
``$\*$-eigen-$\*$ equation'', locally, i.e. away from the walls
\cite{KW}. Here we show that this substitute equation can be
written in a very simple form, even in the presence of an
additional, arbitrary, but regular potential. The more general
applicability of the $\*$-eigen-$\*$ equation is then
demonstrated. First, using an idea from \cite{FM}, we extend it to
a dynamical equation describing time evolution. We then show that
also for general contact interactions,  the $\*$-eigen-$\*$
equation is satisfied locally. Specifically, we treat the most
general possible (Robin) boundary conditions at an infinite wall,
general one-dimensional point interactions, and a finite potential
jump. Finally, we examine a smooth potential, that has simple but
different expressions for $x$ positive and negative. We find that
the \ssee\ is again satisfied locally. It seems, therefore, that
the \ssee\ is generally relevant to the matching of Wigner
functions; it can be solved piece-wise and its solutions then
matched.

\end{abstract}

\vfill\eject
\section{Introduction}

Consider the quantum mechanics of a single particle moving in one
dimension. Infinite potential walls are handled easily in operator
quantum mechanics. The physical wave functions are solutions of
the Schr{\"o}dinger equation away from the walls, and they are
simply required to satisfy boundary conditions at the wall
locations.

In Wigner-Weyl-Moyal quantum mechanics (or deformation
quantization),\footnote{For elementary introductions, see
\cite{DQel}; for more advanced reviews, consult \cite{DQrev}.}
however, infinite potential walls are surprisingly tricky
\cite{DP}. Quantum states are described by Wigner functions, the
Wigner-Weyl transforms (or symbols) of the corresponding density
matrices. Normally, the Wigner functions $\rho(x,p)$ describing
stationary states obey the $\*$-eigen equations (or so-called
``star-genvalue'' equations) \ben H(x,p)\* \rho(x,p)\ =\
\rho(x,p)\* H(x,p)\ =\ E\,\rho(x,p)\ . \label{sgval}\een Here
$H(x,p)$ is the classical Hamiltonian, and \ben \*\ =\
\exp\left\{\frac{i\hbar}{2}\left(\lpx\rpp -
\lpp\rpx\right)\right\}  \label{stprod}\een is the
Gr{\"o}newold-Moyal star product. In the presence of an infinite
potential wall, the symbol of the well-known density matrix does
not satisfy the $\*$-eigen equation.

What goes wrong? One possibility is that the symbol of the density
matrix is not physical. It was shown that this is not the case,
however, in \cite{KW} and \cite{KWi}. In \cite{KW}, the infinite
potential wall was treated as a limit of an exponential
(Liouville) potential, for which the $\*$-eigen equation is
solvable \cite{CFZ}. In the $\alpha\to\infinity$ limit  of the
potential $V_0\,e^{2\alpha x}$, its Wigner function approaches the
canonical one, the symbol of the usual density matrix. The details
were spelled out in \cite{KWi}.

Now, $\alpha$ determines both the height and the ``size'' (or
width) $1/\alpha$ of the potential $V_0\,e^{2\alpha x}$. We
believe that it is the zero-size $1/\alpha\to 0$ limit that is the
relevant one here. As just described, if the zero-size limit is
taken after the $\hbar\to 0$ limit, the physical result is
obtained.

Deformation quantization treats quantum mechanics as an
$\hbar$-deformation of ordinary classical mechanics. It therefore
assumes that the usual canonical, classical mechanics is recovered
in the $\hbar\to 0$ limit. Since an infinite potential wall (or
any potential with a zero-size feature) is to be understood as the
zero-size limit of a smooth potential, we would like to take that
limit first, and then $\hbar$-deform, to get its phase-space
quantum mechanics.

The problem arises because the limits $\hbar\to 0$ and
$1/\alpha\to 0$ do not commute. Put another way, no finite de
Broglie wavelength can be considered small relative to the width
of a sharp, infinite potential wall \cite{BDJ}. If the zero-size
limit ($1/\alpha\to 0$) is taken first, classical mechanics is not
retrieved in the $\hbar\to 0$ limit.

Certain phenomena make that clear. For example, there are
non-Newtonian, or para-classical, reflections present in the
$\hbar\to 0$ limit, that are not described by classical mechanics
\cite{BDJ}. To describe them, perhaps some para-classical
mechanics\footnote{Para-classical reflections have been
incorporated into a path-integral formulation in \cite{Pu}.} could
be deformed, but not normal classical mechanics.\footnote{These
para-classical reflections are present in the operator formulation
of quantum mechanics. That formulation does not seem to have
difficulty with infinite potential walls, however; it does not
describe quantum mechanics as a deformation of classical
mechanics.}

The situation is even worse, however. Deformation quantization
treats quantum mechanics as an $\hbar$-deformation of a specific
treatment of ordinary classical mechanics: the canonical,
phase-space formulation. To the best of our knowledge, no such
formulation exists for a system with a zero-size potential
feature, such as the infinite potential wall.\footnote{Of course,
such systems can be described as zero-size limits of ones with
regular potentials.}

How can such potentials be treated in pure deformation
quantization, i.e., in deformation quantization, considered as an
autonomous formulation of quantum mechanics?

Dias and Prata introduced a boundary potential to cure the problem
\cite{DP}. Consistent with the arguments above, the potential is
proportional to $\hbar^2$, and so describes non-classical effects.
The additional boundary term is ad hoc, however. The original
motivation for the work reported in \cite{KW} was to derive this
term from first principles. This has not yet been
achieved.\footnote{See \cite{KWi}, however. It was pointed out
there that even in the operator formulation, the free Hamiltonian
on the half-line must be extended to a self-adjoint operator by
adding a boundary potential at the wall's location. However, the
additional boundary potential is not of the precise form proposed
by Dias and Prata \cite{DP}.}

Instead, in \cite{KW} an alternative method of pure deformation
quantization was found for systems with such infinite-wall
potentials. Most importantly, the result was derived, rather than
postulated. For the infinite wall potentials, the Wigner function
was shown to satisfy a higher-order $\*$-equation locally, i.e.,
away from the walls.  The physical Wigner function can be found by
solving this new equation, then imposing boundary conditions
\cite{DPcomment}.

If the potential energy consists only of infinite potential walls,
the Wigner function was shown to obey the higher-order
$\*$-equation \ben (p^2-E)\*\rho(x,p)\*(p^2-E)\ =\ 0\
\label{twop2}\een locally, i.e. away from those walls. For
simplicity, we use units such that $2m=1$, so that the Hamiltonian
away from the wall location is just $H=p^2$. The new
``$\*$-eigen-$\*$ equation'' is therefore, in this case, just \ben
\big(\, H(x,p)-E \,\big)\,\*\,\rho(x,p)\,\*\,\big(\, H(x,p)-E
\,\big)\ =\ 0\ .\label{starstar}\een

In \cite{KW}, the $\*$-eigen-$\*$ equation equation was shown to
work for various potentials composed of infinite walls, and wells.
It was also generalized to the case when an additional, arbitrary
but regular potential is present. The resulting equation had a
complicated form -- see (\ref{startwoV}) below. Our first result
here is that this complicated expression reduces to the simple
equation (\ref{starstar}), which is therefore completely general.
The demonstration can be found in the next section.

The $\*$-eigen-$\*$ equation is therefore simpler than was
previously thought. This work will also show that it is more
generally applicable than was realized.

It should be mentioned that a comparison of the Dias-Prata
boundary potential solution \cite{DP} and the $\*$-eigen-$\*$
method \cite{KW} was made in \cite{DPcomment}. A certain
equivalence of the two methods was shown, provided suitable
boundary and kinematical conditions are imposed. One difference
was also pointed out, however. It seems reasonably straightforward
to study dynamics using the Dias-Prata boundary potential -- the
equation of motion of the Wigner function is the usual evolution
equation, but with the boundary potential term included. Time
evolution seems problematic using the $\*$-eigen-$\*$ equation,
however. We address this concern in section 3. Borrowing an idea
from \cite{FM}, we show how an equation of motion can be written
that reduces to the $\*$-eigen-$\*$ equation for stationary Wigner
functions.

Section 4 is motivated by the arguments above concerning zero-size
features of potentials. We examine one-dimensional systems with
contact interactions,\footnote{This is the terminology used in
\cite{FCT}.} i.e., point interactions, or sharp reflecting
boundaries. In subsection 4.1, the general, Robin boundary
conditions for the half-line are considered, and the general point
interaction is treated in subsection 4.2. In all cases, the
$\*$-eigen equation is not satisfied locally, while the
$\*$-eigen-$\*$ equation is, just as for the infinite potential
wall. Perhaps most significantly, the {\it finite} potential wall
demonstrates the same behaviour, as shown in section 4.3.

In section 5 we consider a simple potential that has no zero-size
features. While it is smooth everywhere, it is written with
different expressions for $x>0$ and $x<0$. Remarkably, the same
phenomenon occurs: the local {\ssee}s are satisfied for $x$
positive and negative (while the local $\*$-eigen equations are
not). It seems, therefore, that the \ssee\ is generally relevant
to the matching of Wigner functions; it can be solved piece-wise
and its solutions then matched.

Section 6 is our conclusion.

\vskip.5cm\section{Simple form of the $\*$-eigen-$\*$ equation}

Consider the Hamiltonian \ben H_{\alpha}\ =\ p^2\ +\ e^{2\alpha x}
\ +\ V(x)\ , \een where $V(x)$ is an arbitrary, regular potential.
In the $\alpha\to\infinity$ limit, an infinite potential wall is
formed at $x=0$, so that motion is restricted to the negative
$x$-axis.

In \cite{KW}, it was shown that in the same limit, the stationary
Wigner function satisfies the rather cumbersome
equation\footnote{Henceforth, we will set $\hbar=1$.} \bea
\frac{1}{16}\,\partial_{x}^4\, \rho(x,p)\ +\
\frac{(p^2+E)}{2}\,\partial_{x}^2\, \rho(x,p)\ +\
(p^4-2Ep+E^2)\rho(x,p)\ \nn \ +\ (p^2-E) \re
\left[V(x)\star\rho(x,p)\right] \ -\ p\,\partial_{x}
\im\left[V(x)\star \rho(x,p)\right] \qq \nn -\
\frac{1}{4}\,\partial_{x}^2 \re \left [V(x)\star\rho(x,p)\right]\
-\ \im \left[V(x)\star{p}\,\partial_{x}\rho(x,p)\right] \qq \nn +\
\im \left\{ V(x)\star \im \left[V(x)\star\rho(x,p)\right] \right\}
\ +\ \re\left\{V(x)\star \re \left[V(x)\star\rho(x,p)\right]
\right\}\ \nn \ +\ \re\left\{V(x)\star \left [\left
(p^2-E-\frac{1}{4}\,\partial_{x}^2\right )\rho(x,p)\right
]\right\}\ =\ 0\ , \label{startwoV}\eea for $x< 0$. Clearly, this
equation reduces to (\ref{twop2}) when $V=0$. We will now show
that (\ref{startwoV}) can be simplified substantially, to
(\ref{starstar}) above.

First, write \cite{FM}\ben f\*g\ =\ (f, g)\ +\ i\,[f,g]\
,\label{starsas}\een where \ben \left(\,f,\,g\,\right)\ :=\
f\,\cos\left\{\frac{\hbar}{2}(\lpx\rpp - \lpp\rpx)\right\}\,g\
,\label{symstar}\een and \ben \left[\,f,\,g\,\right]\ :=\
f\,\sin\left\{\frac{\hbar}{2}(\lpx\rpp - \lpp\rpx)\right\}\,g\
,\label{asymstar}\een so that \ben (f,g)\ =\ (g,f)\ ,\ \ [f,g]\ =\
-[g,f]\ .\label{sympm}\een  If $f$ and $g$ are both real, then so
are $(f,g)$ and $[f,g]$, and we also have \ben \re(f\* g)\ =\
(f,g)\ ,\ \ \ \im(f\* g)\ =\ [f,g]\ . \label{reimstar}\een
Finally, the following identities  \bea [[f,g],h]\ +\ [[h,f],g]\
+\ [[g,h],f]\ =\ & 0\ \nn\ [(f,g),h]\ +\ ([h,f],g)\ +\ ([h,g],f)\
=\ & 0\ \nn\ [(f,g),h]\ +\ [(h,f),g]\ +\ [(g,h),f]\ =\ & 0\
\label{sJac}\eea ensure the associativity of the star product.

Now, $V=0$ in (\ref{startwoV}) must result in the equation
(\ref{twop2}). Therefore, the first line of (\ref{startwoV}) can
be rewritten as $(p^2-E)\*\rho\*(p^2-E)$, as can be verified
directly. Also notice that \bea & -p\,\partial_x\,f(x,p)\ =\
\im\left(\, p^2\* f(x,p) \,\right)\ =\ \left[p^2,f(x,p) \right]\
\nn\ &(p^2-\frac 1 4\partial_x^2)f(x,p)\ =\ \re\left(p^2\*f(x,p)
\right)\ =\ \left( p^2,f(x,p) \right)\ ,\label{imp2}\eea for real
$f$. Equation (\ref{startwoV}) can therefore be simplified to \bea
(p^2-E)\*\rho\*(p^2-E)\ \qq\QQ\nn \ +\ \left(\, p^2-E, \left(
V,\rho\right) \,\right) \ +\ \left[\, p^2-E, \left[ V, \rho
\right] \,\right] \ +\ \left[\, V, [ p^2-E, \rho ] \,\right] \qq
\nn +\ \left[\, V, \left[ V, \rho \right] \,\right] \ +\ \left(\,
V, \left( V, \rho \right) \,\right)\  +\ \left(\, V, \left( p^2-E,
\rho \right) \,\right)\ =\ 0\ . \label{startwoVi}\eea Here we have
also used that $[E,f]=0$ for any $f$, since $E$ is a constant. It
is then a simple matter to verify that (\ref{startwoVi}) reduces
to \ben \left(\, H-E, \left(\, H-E, \rho \,\right) \,\right)\ +\
\left[\, H-E, \left[\, H-E, \rho \,\right] \,\right]\ =\ 0\
.\label{ststimre}\een Here we have defined the Hamiltonian away
from the wall as \ben H\ =\ p^2\ +\ V(x)\ .\label{Haw}\een But
(\ref{ststimre}) is just (\ref{starstar}), the \ssee.

It is also interesting to notice that the \ssee\ can be written as
\ben (H-E)\,\bar{\*}\,\big(\, (H-E)\*\rho \,\big)\ =\
0.\label{stbst}\een Here $\bar\* = \exp[-i(\lpx\rpp-\lpp\rpx)/2]$
is the complex conjugate of $\*$.

\vskip.5cm\section{Time evolution and the $\*$-eigen-$\*$
equation}

So far we have only considered the Wigner functions \ben
\rho(x,p)\ =\ \frac{1}{\pi}\,\int_{-\infinity}^\infinity dy\,
e^{-2ipy}\, \psi(x+y)\,\bar\psi(x-y)\ \label{WWpure}\een derived
from the density matrices for a stationary energy eigenstate.
These Wigner functions have no explicit dependence on time. We now
want to study the time dependence of Wigner functions. Let us
denote by $R(x,p;t)$ the symbol of a density matrix element that
has explicit time dependence, in order to distinguish it from one
with none.

It is the time-independent wave function $\psi(x)$ that enters
(\ref{WWpure}). Simply introducing its time dependence,
$\Psi(x,t):= \psi(x)e^{-iEt}$, has no effect. We must therefore
consider Wigner functions that are the symbols of off-diagonal
density matrix elements in the stationary state basis: \bea
R_{12}(x,p;t)\ =\ \frac{1}{\pi}\,\int_{-\infinity}^\infinity dy\,
e^{-2ipy}\, \Psi_1(x+y,t)\,\bar\Psi_2(x-y,t)\ \nn \ =\
\frac{1}{\pi}\,\int_{-\infinity}^\infinity dy\, e^{-2ipy}\,
\psi_1(x+y,t)e^{-iE_1t}\,\bar\psi_2(x-y)e^{iE_2t}\ .
\label{WWoff}\eea If the wave functions satisfy the Schr\"odinger
equation, then we have \ben \frac{\partial R_{12}}{\partial t}\ =\
\frac 1 {i\hbar}\,\left( H\* R_{12} - R_{12}\* H \right)\ =\ \frac
1 {i\hbar}\, [H, R_{12}]\ ,\label{tdep}\een the fundamental
dynamical equation.

From the pure deformation quantization point of view, (\ref{tdep})
must be the starting point. So, how are the $\*$-eigen equations
derived from it? Substituting the ansatz \ben R_{12}(x,p;t) =
\rho_{12}(x,p)\,\exp[-i(E_1-E_2)t]\label{Rrtans}\een yields \ben
[H, \rho_{12}]\ =\ (E_1-E_2)\,\rho_{12}\ .\label{sinHE}\een The
$\*$-eigen equations are therefore obtained if \ben (H,
\rho_{12})\ =\ (E_1+E_2)\,\rho_{12}\ \label{cosHE}\een can also be
derived.

In \cite{FM}, this was done by introducing a complex time
$t\rightarrow z:=t-is$. Here we will use the same trick to find a
dynamical equation that corresponds to the \ssee\
(\ref{starstar}), or the generalization thereof: \ben \left[\,
H(x,p)-E_1 \,\right]\,\*\,\rho_{12}(x,p)\,\*\,\left[\, H(x,p)-E_2
\,\right]\ =\ 0\ . \label{ststi.ii}\een To that end, we write \ben
\tilde R_{12}(x,p,z)\ :=\
\frac{1}{\pi}\,\int_{-\infinity}^\infinity dy\, e^{-2ipy}\,
\psi_1(x+y)e^{-iE_1z}\,\bar\psi_2(x-y)e^{iE_2\bar z}\ ,
\label{WWoffz}\een and find \bea H(x,p)\*\tilde R_{12}(x,p,z)\ =\
H(x+\frac i 2\rpp, p-\frac i 2\rpx)\,\tilde
R_{12}(x,p,z)\qq\quad\nn  =\ c_{12}\, \int_{-\infinity}^\infinity
dy\, e^{-2ipy}\, H(x+y, p-\frac i 2\rpx)\,
\psi_1(x+y)\bar\psi_2(x-y)\quad\ \ \ \  \nn \   =\ c_{12}\,
\int_{-\infinity}^\infinity dy\, e^{-2ipy}\, H(x+y, \frac i
2\lpy-\frac i 2\rpx)\, \psi_1(x+y)\bar\psi_2(x-y)\ ,\ \
 \label{HsrEi}\eea where we have set \ben c_{12}\ :=\ \exp[-i(E_1z-E_2\bar
z)]/\pi\ \label{ciii}\een to save writing. Integrating by parts,
this becomes \bea c_{12}\, \int_{-\infinity}^\infinity dy\,
e^{-2ipy}\, H(x+y, -\frac i 2\rpy-\frac i 2\rpx)\,
\psi_1(x+y)\bar\psi_2(x-y)\nn \  =\ E_1\,\tilde R_{12}(x,p,z)\
 .\qq \label{HsrEii}\eea Similarly, we can show that \ben
 \tilde R_{12}(x,p,z)\* H(x,p)\ =\ \tilde R_{12}(x,p,z)\,
 E_2\ .\label{rsHE}\een Consequently, the dynamical equation \ben
 \left(\, i\frac{\partial}{\partial z}-H
 \,\right)\,\*\,\tilde R_{12}(x,p,z)\,\*\,\left(\, -i\frac{\partial}{\partial{\bar z}}-H
 \,\right)\ =\ 0
 \label{dzzbE}\een is obeyed.

 If we substitute \ben
\tilde R_{12}(x,p,z) = \rho_{12}(x,p)\,\exp[-i(E_1z-E_2\bar z)]\
,\label{Rrzans}\een then \ben
 \left(\, E_1-H
 \,\right)\,\*\,\rho_{12}\,\*\,\left(\, E_2-H
 \,\right)\ =\ 0
 \label{sszE}\een follows. As a special case, when $\psi_1=\psi_2$,
 the \ssee\ (\ref{starstar}) is retrieved. These equations can be
 solved to recover the physical time evolution of the density
 matrix symbols.

 Note that if $z=\bar z$, then the \ssee\ (\ref{starstar}) for the
 ``diagonal'' Wigner
 function could not have been derived in this way. The
 complexification of time used here is therefore necessary,
 if somewhat artificial, as
 well as having appeared before in \cite{FM}.

\vskip.5cm\section{Potentials with zero-size features: contact
interactions}

The main point of this article is that the \ssee\ applies more
generally. In particular, it applies to more general potentials.
Included are those with zero-size features, describing the
so-called contact interactions, considered in this section.

\vskip.5cm\subsection{Robin boundary conditions on the half line}

As our first example, consider an infinite potential wall at $x=0$
that prevents motion on the positive $x$-axis. Equivalently, we
can consider motion restricted to the half-line $x<0$, with a
point interaction at $x=0$ that is necessary for a self-adjoint
Hamiltonian. The point interaction has the effect of imposing
boundary conditions on the wave function.

The most general point interaction at $x=0$ results in the most
general, mixed or Robin, boundary conditions \ben \psi(0)\ +\
L\,\psi'(0)\ =\ 0\ \label{Robpsi}\een on the wave function
$\psi(x)$. Here $\psi'=d\psi/dx$ and $L$ is a real length scale.

A (scattering) wave function satisfying these boundary conditions
is \cite{FCT}  \ben \psi (x)\ \propto\
e^{ikx}+e^{i\delta_k}e^{-ikx}\ ,\label{psiL}\een for $x\le 0$, if
\ben kL\ =\ \cot(\delta_k/2)\ .\label{deltaL}\een  Notice that
$\delta_{-k}\ =\ -\delta_k\ .$ These Robin boundary conditions
interpolate between the standard Dirichlet ones at $L=0$, and the
Neumann boundary conditions for $L=\infinity$. In these cases,
$\delta_k=\pi$ and $0$, respectively.

The Wigner function derived from this wave function by a
Weyl-Wigner transform is, for $x<0$,  \ben \rho \ =\
\frac{\sin[2(p-k)x]}{(p-k)}\ +\ \frac{\sin[2(p+k)x]}{(p+k)}\ +\
2\cos(2kx-\delta_k)\,\frac{\sin(2px)}{p}\ .\label{rhoL}\een It
satisfies $\rho (x,-p) = \rho (x,+p)$, and $\rho (0,p)=0$.

Clearly, this Wigner function is real, and depends on $x$. As a
consequence, we know it cannot satisfy the $\*$-eigen equations
(\ref{sgval}) away from $x=0$. By (\ref{reimstar}), the $\*$-eigen
equations imply \ben  [p^2, \rho]\ =\ p\,\partial_x\rho\ =\ 0\ ,
\label{imstareigen}\een for $\rho = \bar\rho$.

In the free case ($V=0$), we can understand the problem in a
slightly deeper way. There is a conflict between the reality of
solutions of the $\*$-eigen equation and the presence of both
momenta $\pm\sqrt{E}$, such as is necessary when reflections
occur. See the Appendix.

The alternative form of the Wigner function \bea \rho \ =&\
\sin[2(p-k)x]/(p-k)\ +\ \sin[2(p+k)x]/(p+k)\ \nn&\ \ +\
\sin[2(p-k)x+\delta_k]/p\ +\ \sin[2(p+k)x-\delta_k]/p\
\label{rhoLi}\eea makes it straightforward to see that the \ssee\
equation is obeyed, however. With $H=p^2$ and $E=:k^2$, it becomes
\ben \left[\, (p^2-k^2)^2+2(p^2+k^2)\,\partial_{(2x)}^2 +
\partial_{(2x)}^4 \,\right]\,\rho\ =\ 0\ .\label{sseepk}\een
But \bea &\left[\, (p^2-k^2)^2+2(p^2+k^2)\,\partial_{(2x)}^2 +
\partial_{(2x)}^4 \,\right]\, \sin[2(p\pm k)x\mp \delta_k]\QQ\nn &\ =\
\left[\, (p^2-k^2)^2 -2(p^2+k^2)(p\pm k)^2 + (p\pm k)^4
\,\right]\, \sin[2(p\pm k)x\mp \delta_k]\qq\nn &\ =\ (p\pm k)^2
\left[\, (p\mp k)^2-2(p^2+k^2) + (p\pm k)^2 \,\right]\,
\sin[2(p\pm k)x\mp \delta_k]\qq\ ,\label{rhoLdem}\eea which
clearly vanishes. The Wigner function (\ref{rhoL}) therefore
satisfies the \ssee.

Incidentally, just as $(p^2-k^2)\*\sin[2(p\pm k)x]\*(p^2-k^2) =
0$, so does $(p^2-k^2)\*\cos[2(p\pm k)x]\*(p^2-k^2) = 0$.
Consequently, the four fundamental solutions of the free \ssee\
can be taken to be \bea &\left\{\, \cos(2kx)\cos(2px),
\cos(2kx)\sin(2px), \sin(2kx)\cos(2px), \sin(2kx)\sin(2px)
\,\right\}\nn &{\rm or}\ \ \left\{\, \cos[2(p+k)x], \cos[2(p-k)x],
\sin[2(p+k)x], \sin[2(p-k)x] \,\right\}\ ,\nn &{\rm or}\ \
\left\{\,\, e^{2i(p+k)x},\, e^{2i(p-k)x},\, e^{-2i(p+k)x},\,
e^{-2i(p-k)x} \,\,\right\}\ .\label{ffour}\eea The third basis can
also be found by realizing that $p^2-k^2 = (p\pm k)\*(p\mp k)$,
and that \ben (p\pm k)\* e^{-2i(p\pm k)x}\ =\ 0\ \ \ \rightarrow\
\ \ e^{2i(p\pm k)x}\* (p\pm k)\ =\ 0\ . \label{expipk}\een Any
linear combination of the four solutions from any of these bases,
with $x$-independent coefficient functions, satisfies the \ssee.

A bound state also exists for $L>0$, with wave function \ben
\psi(x)\ =\ \theta(-x)\, \sqrt{\frac 2 L}\,\, e^{x/L}\ ,
\label{Rbdwf}\een of energy $E=-1/L^2$.

Below we will need to consider the more general situation when
$H-E=p^2+\kappa^2$, with $\kappa^2>0$. Now $p^2+\kappa^2 = (p\pm
i\kappa)\*(p\mp i\kappa)$, and \ben (p\pm i\kappa)\*
e^{-2ipx}e^{\pm 2\kappa x}\ =\ 0\ \ \ \rightarrow\ \ \
e^{2ipx}e^{\pm 2\kappa x}\* (p\mp i\kappa)\ =\ 0\ .
\label{exppk}\een Therefore, any of \ben \left\{\,\, e^{\pm
2ipx}\,e^{\pm'2\kappa x} \,\,\right\}\ \label{fourneg}\een
satisfies the \ssee, as does any linear combination with
$x$-independent coefficient functions.

The Wigner-Weyl transform of the wave function (\ref{Rbdwf})
produces the Wigner function \ben \pi\rho[\psi]\ =\ -\frac 2
L\,\frac{1}{p}\,\, \sin(2px)\, e^{2x/L}\ ,  \label{Rbdwig}\een for
$x<0$, while $\rho$ vanishes for $x>0$. Identifying $\kappa=1/L$,
then, by (\ref{fourneg}), the \ssee\ is satisfied away from the
wall at $x=0$.

\vskip.5cm\subsection{General point interaction on the real line}

A four-parameter family of point interactions exists in
one-dimensional quantum mechanics (see \cite{CFT} and \cite{CNF},
e.g.). The corresponding potentials include the Dirac
delta-function, that has the effect of relating the values of the
wave function and its derivative on its two sides. The general
point interaction imposes more general matching conditions
involving the wave function and its $x$-derivatives on its sides.

With a point interaction at $x=0$, and wave function \ben \psi(x)\
=\ \theta(-x)\,\psi_-(x)\ +\ \theta(x)\,\psi_+(x)\ ,
\label{wfpm}\een the general matching conditions are \bea
-\psi_+'(0)-\alpha\psi_-'(0)\ =\ \beta\psi_-(0)\nn
-\delta\psi_-'(0) - \gamma\psi_-(0)\ =\ \psi_+(0)\ .
\label{genpt}\eea Here $\psi'$ indicates the space derivative of
the wave function,  and the real parameters
$\alpha,\beta,\gamma,\delta$ are related by \ben \alpha\gamma\ -\
\beta\delta\ =\ 1\ . \label{pardet}\een

For any wave function of the form (\ref{wfpm}),  the Wigner
function is \bea \pi\,\rho[\psi]\ =\ \theta(-x)\int_x^{-x}dy\,
e^{-2ipy}\, \psi_-(x+y) \,\bar\psi_-(x-y)\nn +\
\int_{-\infinity}^{-|x|}dy\, e^{-2ipy}\, \psi_-(x+y)
\,\bar\psi_+(x-y) \quad\nn +\ \int^{\infinity}_{|x|}dy\,
e^{-2ipy}\, \psi_+(x+y) \,\bar\psi_-(x-y)\ \quad\nn \ +\
\theta(x)\int_{-x}^{x}dy\, e^{-2ipy}\, \psi_+(x+y)
\,\bar\psi_+(x-y)\ . \label{wigpm}\eea

Consider first the bound state for a point interaction at $x=0$,
of energy $E=-\kappa^2$, present if \ben \beta+\delta
\kappa^2+\kappa(\alpha+\gamma)\ =\ 0\, ,\ {\rm with}\ \kappa>0\
.\label{boundgpi}\een The wave function is \ben \psi_\pm(x)\ =\
\psi_\pm(0)\,e^{\mp\kappa x}\ , \label{psipmb}\een with \ben
\frac{\psi_+(0)}{\psi_-(0)}\ =\ \alpha+ \frac\beta\kappa\ =\
-\gamma-\delta\kappa\ .\label{bdmatch}\een A simple calculation
gives \bea \pi\,\rho[\psi]\ =\ -\ \frac{e^{2\kappa
x}\,e^{2ipx}}{2i}\, \bar\psi_-(0)\, \left\{\, \frac{\psi_-(0)}{p}\
-\ \frac{\psi_+(0)}{p-i\kappa} \,\right\}\nn \quad +\
\frac{e^{2\kappa x}\,e^{-2ipx}}{2i}\, \psi_-(0)\, \left\{\,
\frac{\bar\psi_-(0)}{p}\ -\ \frac{\bar\psi_+(0)}{p+i\kappa}
\,\right\}\ ,\label{wbdneg} \eea for $x<0$, and \bea
\pi\,\rho[\psi]\ =\ \ \frac{e^{-2\kappa x}\,e^{2ipx}}{2i}\,
\bar\psi_+(0)\, \left\{\, \frac{\psi_+(0)}{p}\ -\
\frac{\psi_-(0)}{p+i\kappa} \,\right\}\nn  \quad -\
\frac{e^{-2\kappa x}\,e^{-2ipx}}{2i}\, \psi_+(0)\, \left\{\,
\frac{\bar\psi_+(0)}{p}\ -\ \frac{\bar\psi_-(0)}{p-i\kappa}
\,\right\}\ ,\label{wbdpos} \eea for $x>0$. In view of
(\ref{fourneg}), we see that the Wigner function satisfies the
\ssee\ away from the point interaction.

Let us now consider the Wigner function related to a scattering
wave function \cite{CNF}\ben \psi(x)\ =\ \theta(-x)\,\left[\,
e^{ikx}\ +\ Re^{-ikx}  \,\right]\ +\ \theta(x)\,T e^{ikx}\
\label{wfsc}\een satisfies the matching conditions (\ref{genpt}),
with the constants \bea T\ =\ -2ik/D\ ,\ \ R\ =\
\left[\beta+\delta k^2+i k(\alpha-\gamma)\right]/D\ ,\nn D\ =\
-\beta+\delta k^2+i k(\alpha+\gamma)\ .\qquad\quad\label{TRD}\eea
The transmission and reflection amplitudes satisfy \ben |R|^2\ +\
|T|^2\ =\ 1\ ,\label{RT1}\een because of (\ref{pardet}). Putting
$\psi_-(x) = e^{ikx}+Re^{-ikx}$ and $\psi_+(x) = Te^{ikx}$ in
(\ref{wigpm}) produces \bea  \pi\, \rho[\psi](x<0)\ =\ -\left\{\,
\frac{\re(1-T)}{p-k} \ +\ \frac{\re(R)}{p} \,\right\}\,
\sin[2(p-k)x]\quad\nn \ -\ \left\{\, \frac{|R|^2}{p+k}\ +\
\frac{\re[(1-T)\bar R]}{p} \,\right\}\, \sin[2(p+k)x]\nn \ -\
\left\{\, \frac{\im(1- T)}{p-k}\  +\ \frac{\im(R)}{p} \,\right\}\,
\cos[2(p-k)x]\nn \ -\ \left\{\, \frac{\im[(1-T)\bar R])}{p}
\,\right\}\, \cos[2(p+k)x]\ , \label{Wiggptxneg}\eea for $x<0$,
and for $x>0$: \bea \pi\,\rho[\psi](x>0)\ =\ -\,\left\{\,
\frac{\re[T(1-\bar T)]}{p-k}\ +\ \frac{\re(T\bar R)}{p}
\,\right\}\,\sin[2(p-k)x]\ \ \nn \ +\ \left\{\,
\frac{\im(T)}{p-k}\ +\ \frac{\im(T\bar R)}{p} \,\right\}\,
\cos[2(p-k)x]\ . \label{Wiggptxpos}\eea Interestingly, these
results can be written in a more compact form as \bea -\pi\,
\rho[\psi](x<0)\ =\   \im\,\Bigg\{\, \bigg( \frac{1-T}{p-k}\ +\
\frac{R}{p} \bigg)\, e^{2i(p-k)x}\qq\nn +\ \bigg(\, \frac{R\bar
R}{p+k}\ +\ \frac{(1-T)\bar R}{p} \,\bigg)\, e^{2i(p+k)x}
\,\Bigg\} \label{Wiggptxnegc}\eea for $x<0$, and for $x>0$: \bea
-\,\pi\,\rho[\psi](x>0)\ =\ \im\,\Bigg\{\, \bigg( \frac{(1-T)\bar
T}{p-k}\ +\ \frac{R\bar T}{p} \,\bigg)\, e^{2i(p-k)x} \,\Bigg\}\ .
\label{Wiggptxposc}\eea These expressions make it clear that the
Wigner functions do not satisfy the $\*$-eigen equations, but do
obey the $\*$-eigen-$\*$ equation.

A subtlety must be discussed here, however. When performing the
integrations for the ``$\pm\mp$ cross terms'' \ben 2\re\,
\int_{|x|}^\infinity dy\, e^{-2ipy}\, \psi_+(x+y)\bar\psi_-(x-y)
\label{crossre}\een of (\ref{wigpm}), we have dropped oscillating
contributions at $y=\pm\infinity$. This could  be implemented
formally by changing $p\to p-i\epsilon$, $0<\epsilon$, in
(\ref{crossre}), then taking the limit $\epsilon\to 0$ after
integrating over $y$.

The use of non-normalizable plane waves as the scattering wave
functions is the origin of the undefined terms that we omitted.
Scattering is certainly treatable with Wigner functions, by using
wave packets \cite{CZ}. Such an analysis is beyond our scope,
however, since we are mainly interested in the time-independent
Wigner functions and their equations of motion.

\vskip.5cm\subsection{Finite potential jump}

If the usefulness of the \ssee\ is tied to the zero-size property
of a feature of the relevant potential, then it should apply to a
finite potential jump, not just to an infinite one. Here we show
that this is indeed the case.

The Wigner function for the potential $V(x) = V_0\,\theta(x)$,
with $E<V_0$, was studied in \cite{LS}.  The relevant wave
function can be expressed as\ben \psi(x)\ \propto\
\theta(-x)\,\cos(kx-\alpha/2)\ +\
\theta(x)\,\cos(\alpha/2)\,e^{-\kappa x}\ ,\label{wfjump}\een
where $E=k^2$ again, and $\kappa^2:=V_0-E$. For continuity of
$d\psi/dx$ at $x=0$, \ben e^{i\alpha}\ =\
\frac{ik+\kappa}{ik-\kappa}\label{alphak}\een is required.

For $x<0$, the corresponding Wigner function is \cite{LS} \bea -\
\frac{k[k(2p-k)+\kappa^2]}{[\kappa^2+(2p-k)^2]4p(p-k)}\,\sin[2(p-k)x]\qq\nn
-\
\frac{k[k(2p+k)-\kappa^2]}{[\kappa^2+(2p+k)^2]4p(p+k)}\,\sin[2(p+k)x]\qq\nn
+\ \frac{\kappa k\,\cos[2(p-k)x]}{2p[\kappa^2 +(2p-k)^2 ]}\ -\
\frac{\kappa k\, \cos[2(p+k)x]}{2p[\kappa^2 +(2p+k)^2 ]}\ ,
\label{njumprho}\eea up to a multiplicative constant. By
(\ref{ffour}), we see that it satisfies the \ssee\ for $H=p^2$,
valid for $x<0$.

The Wigner function is proportional to \bea k^2\,e^{-2\kappa
x}\,\, \bigg\{\
\frac{4\kappa}{[(2p+k)^2+\kappa^2]\,[(2p-k)^2+\kappa^2]}\,
\cos(2px)\qq\nn +\ \frac{(k^2+\kappa^2-4p^2)}{
[(2p+k)^2+\kappa^2]\,2p\,[(2p-k)^2+\kappa^2]}\sin(2px)\ \bigg\}\
.\label{pjumprho}\eea for $x>0$, where  $H-E=p^2+\kappa^2$. The
\ssee\ therefore becomes \ben \left[\,
(p^2+\kappa^2)^2+2(p^2-\kappa^2)\,\partial_{(2x)}^2 +
\partial_{(2x)}^4 \,\right]\,\rho\ =\ 0\ .\label{sseepka}\een
But \bea &\left[\,
(p^2+\kappa^2)^2+2(p^2-\kappa^2)\,\partial_{(2x)}^2 +
\partial_{(2x)}^4 \,\right]\, e^{-2\kappa x}\, e^{\mp 2ipx}\QQ\nn
&\ =\ (p\mp i\kappa)^2 \left[\, (p\mp i\kappa)^2-2(p^2-\kappa^2) +
(p\pm i\kappa)^2 \,\right]\, e^{-2\kappa x}\, e^{\mp 2ipx}\ =\ 0\
.\label{rhojdem}\eea We conclude that the Wigner function
(\ref{pjumprho}) satisfies the \ssee\ for $x>0$ as well. Clearly,
(\ref{rhojdem}) amounts to an explicit confirmation of
(\ref{fourneg}).

For $E>V_0$, we have the extended wave function \ben \psi_-(x)\ =\
e^{ikx}\ +\ R\,e^{-ikx}\ ,\ \ \psi_+(x)\ =\ T\,e^{i\ell x}\ ,
\label{extwfjump}\een where $k^2:=E$, and $\ell^2:=E-V_0$. As
discussed in the last section, we use \bea \pi\,\rho[\psi]\ =\
\theta(-x)\int_x^{-x}dy\, e^{-2ipy}\, \psi_-(x+y)
\,\bar\psi_-(x-y)\nn +\ \lim_{\ \ \epsilon\to 0_+\,}2\,\re\,
\int_{|x|}^\infinity dy\, e^{-2i(p-i\epsilon) y}\,
\psi_+(x+y)\bar\psi_-(x-y)\nn \ +\ \theta(x)\int_{-x}^{x}dy\,
e^{-2ipy}\, \psi_+(x+y) \,\bar\psi_+(x-y)\ . \label{wigpmii}\eea

For $x>0$, this gives \ben \pi\,\rho[\psi]\ =\ \im\,\bigg\{\,
e^{-2ix(p-\ell)}\, \left[\, -\ \frac{|T|^2}{p-\ell}\ +\ \frac{2
T\bar R}{2p+k-\ell}\ +\ \frac{2T}{2p-k-\ell} \,\right]\,\bigg\}\ ;
\label{evzpos}\een and for $x<0$, \bea \pi\,\rho[\psi]\ =\
\im\,\bigg\{\, e^{-2ix(p-k)}\, \left[\, \frac{1}{p-k}\ -\
\frac{2\bar T}{2p-k-\ell}\ +\ \frac{\bar R}{p} \,\right]\nn +\
e^{-2ix(p+k)}\, \left[\, \frac{|R|^2}{p+k}\ -\ \frac{2\bar T
R}{2p+k-\ell}\ +\ \frac{R}{p} \,\right] \,\bigg\}\ .
\label{evzneg}\eea Putting $\ell=k$ in these formulas yields the
results (\ref{Wiggptxposc}) and (\ref{Wiggptxnegc}), as should be.
Again, it is clear that while these real and $x$-dependent Wigner
functions cannot obey the $\*$-eigen equations, they do satisfy
the $\*$-eigen-$\*$ equation.

Incidentally, notice that our conclusions are similar for the two
cases $E<V_0$ and $E>V_0$. Interestingly, no non-Newtonian
(para-classical) reflections occur for $E<0$, but they do for
$E>V_0$.

\vskip.5cm\section{Matching}

So far, our results seem to support the hypothesis that it is the
zero-size feature of potentials that are necessary for the \ssee\
to be useful.

However, consider a potential that can only be written simply in
different regions of the $x$-axis, but is smooth and does not have
any physical characteristics at the points separating those
regions. As a simple example, we will treat the classical
Hamiltonian \ben H\ =\ p^2\ +\ \theta(x)\,x^2\ ,
\label{zerosho}\een describing a free particle for $x<0$, and a
harmonic force acting when $x>0$. The potential $V(x) =
\theta(x)\,x^2$ has no special properties at $x=0$ -- it is
completely smooth there, for example.

In Schr\"odinger quantum mechanics, one would just solve the
Schr\"odinger equation locally, i.e., in each region, and then
match the solutions. Even contact interactions simply impose
matching conditions on the wave functions across the zero-size
features of the potential. The systems we have considered so far
make it clear that na\"ive matching does not also work for quantum
mechanics in phase space: one cannot obtain the correct Wigner
function by solving the $\*$-eigen equation in different regions,
and then matching them. Is the \ssee\ still relevant even in the
absence of zero-size potential features?

A simple stationary state for this system of Hamiltonian
(\ref{zerosho}), of energy $E=1$, has a wave function of the form
(\ref{wfpm}) with \ben \psi_-(x)\ =\ \cos(x)\ ,\ \ {\rm and}\ \ \
\psi_+(x)\ =\ e^{-x^2/2}\ .\label{psipmmatch}\een The
corresponding Wigner function \ben
 \rho(x,p)\ =\ \theta(-x)\,\rho_-(x,p)\ +\ \theta(x)\,\rho_+(x,p)
 \label{wignerpm}\een can be calculated using (\ref{wigpm}), or
 (\ref{wigpmii}). We find \bea  4\,\rho_-(x,p)\ =\
 \cos[2(p-1)x]\,\left\{\,
 2\sqrt{2\pi}\,e^{-(2p-1)^2/2} \,\right\} \QQ\nn \qq +\
 \cos[2(p+1)x]\,\left\{\,
 2\sqrt{2\pi}\,e^{-(2p+1)^2/2} \,\right\} \QQ\nn -\
 \sin[2(p-1)x]\,\left\{\, \frac 1{p-1}\, +\, \frac 1 p\, +\,
 2i\sqrt{2\pi}\,\,\erf\left[
 \frac{i(2p-1)}{\sqrt{2}}
 \right]\,e^{-(2p-1)^2/2} \,\right\}
 \nn -\
 \sin[2(p+1)x]\,\left\{\, \frac 1{p+1}\, +\, \frac 1 p\, +\,
 2i\sqrt{2\pi}\,\,\erf\left[
 \frac{i(2p+1)}{\sqrt{2}}
 \right]\,e^{-(2p+1)^2/2} \,\right\}\, .\label{rhommatch}\eea
 This is a real function that depends on $x$. It therefore does
 not satisfy (the imaginary parts of) the free $\*$-eigen equations.
 Since its
 $x$-dependence is described by a linear combination of the
 functions of (\ref{ffour}), however, it does satisfy
 the free \ssee.

So, the phenomenon of satisfying the local \ssee\ instead of the
local $\*$-eigen equation, has nothing to do with the special
features (zero-size features, e.g.) of potentials. Nothing unusual
is described by (\ref{zerosho}) at $x=0$.

The Wigner function for $x>0$ is \bea {2\sqrt{2\pi}}\,\rho_+(x,p)\
=\
 e^{2ix(p+1)-(2p+1)^2/2}\ \, \erfc \big[ \sqrt{2}x+i(2p+1)/\sqrt{2} \big] \qq\nn +\
 e^{-2ix(p+1)-(2p+1)^2/2}\ \, \erfc \big[ \sqrt{2}x-i(2p+1)/\sqrt{2} \big] \qq\nn
 -\
 e^{2ix(p-1)-(2p-1)^2/2}\ \, \erfc \big[ \sqrt{2}x+i(2p-1)/\sqrt{2} \big] \qq\nn
 -\
 e^{-2ix(p-1)-(2p-1)^2/2}\ \, \erfc \big[ \sqrt{2}x-i(2p-1)/\sqrt{2} \big]\ \qq\nn
 +\ \sqrt{2}\, e^{-x^2-p^2}\,\,\left\{\,
 \erf(x+ip)\, +\,
 \erf(x-ip)\,\right\} \ .\qq
 \label{rhopmatch}\eea

 For the local Hamiltonian, $H_+=p^2+x^2$, the imaginary parts of the
 $\*$-eigen equations require \ben (p\,\partial_x
 - x\,\partial_p) \,\rho(x,p)\ =\ 0\ , \label{imshose}\een
 implying that the solution should be a function only of $H_+$, not
 of both $x$ and $p$, independently. Clearly, then, (\ref{rhopmatch}) does not
 satisfy the local $\*$-eigen equation.

 With $H\to H_+=x^2+p^2$, however, the local ($x>0$) \ssee\ is
 \bea 0\ =\ \big[\, (x^2+p^2-E)^2 - 1 \,\big]\,\rho\
-\ 2x\,\partial_x\rho\ -\ 2p\,\partial_p\rho\ \qq\qq\nn \ -\
(x^2+p^2-E)(\partial_x^2+\partial_p^2)\rho/2\ +\
x^2\partial_p^2\rho\ -\ 2xp\,\partial_x\partial_p\rho\ +\
p^2\partial_x^2\rho\nn \ +\ (\partial_x^2 +
\partial_p^2)^2\rho/16\ .\QQ\label{sseepmatch}\eea
With $E=1$, the local Wigner function (\ref{rhopmatch}) satisfies
this equation.

These results are consistent with the following conjecture.
Suppose \ben H\ =\ \theta(-x)\,H_-\ +\ \theta(x)\, H_+\
,\label{hampm}\een so that the Wigner function has the form
(\ref{wignerpm}). Then, although $(H_\pm-E)\*\rho_\pm$ does not
necessarily vanish, we still have \ben (H_\pm-E)\*\rho(x,p) \*
(H_\pm-E)\ =\ 0\ . \label{pmconjecture}\een

\vskip.5cm\section{Conclusion}

In \cite{KW}, an alternative approach to the pure deformation
quantization of systems with infinite potential walls was derived
from first principles. Here we have demonstrated that its
substitute equation, the \ssee, is much simpler than was
previously thought, taking the form (\ref{starstar}) for
potentials constructed from infinite walls/wells, with an
arbitrary, but regular, additional potential.

Not only is the \ssee\ simple, it is also more generally
applicable than was realized. For one thing, we showed here it can
be modified to describe the time evolution of such systems (see
section 3). More importantly, we demonstrated its applicability to
other contact interactions, i.e. to other potentials with
zero-size features. The one-dimensional systems we treated here
are: infinite walls with Robin boundary conditions, general point
interactions, and, most significantly, finite potential jumps. For
all these cases, the local $\*$-eigen equations are not satisfied,
while the local \ssee\ is.

Most revealingly, the same properties were shown to apply to
system with a potential written differently for two adjacent
regions of configuration space, as in  (\ref{zerosho}).
Specifically, the particle experiences different local
Hamiltonians for $x<0$ and $x>0$, and while the corresponding
local $\*$-eigen equation is not obeyed, the local \ssee\ is.

What is common to the final example and the contact interactions
is that the Hamiltonian (or potential) is most easily viewed
piece-wise, using local expression in each region. In operator
quantum mechanics, the wave functions would be found in each
region, i.e. locally, and then matched at the boundaries
separating the regions.

It seems, therefore, that the \ssee\ is generally relevant to the
matching of Wigner functions. We conjecture that it can be solved
locally, i.e. piece-wise, and its solutions then matched to get
the ``global'', physical Wigner function.

\vskip1cm \noindent{\bf Acknowledgments}\hfill\break I thank
Sergei Kryukov for collaboration in the early stages of this work.
For comments, I am grateful to him and to Terry Gannon. This
research was supported in part by NSERC of Canada.


\vskip1cm
\noindent{\large{\bf Appendix:\ \  Free \ssee}} 

For the free Hamiltonian $H=p^2$, putting $E:=k^2$, \ben H\ -\ E\
=\ (p-k)\*(p+k)\ =\ (p+k)\*(p-k)\ .
 \label{pkstpk}\een So (using the associativity of the $\*$-product)
 $f_{\pm k}(x,p):=\exp[-2ix(p\mp k)]$
 satisfy the \ssee, since \ben (p\mp k)\* f_{\pm k}\ =\ 0\ =\ \bar
 f_{\pm k}\*(p\mp k)\ .\label{pkf}\een
 If reflections occur, then both components $f_k$ and $f_{-k}$
 must be present.

 Equations (\ref{pkf}) imply that any linear
 combination of $f_k$ and $f_{-k}$ satisfies the $\*$-eigen
 equation, \bea (p^2-k^2)\*\big( \alpha_k\,f_k\ +\ \alpha_{-k}\,f_{-k}
 \big)\QQ\QQ\nn \ =\ \alpha_k\,(p+k)\*(p-k)\* f_k\ +\
 \alpha_{-k}\,(p-k)\*(p+k)\*f_{-k}\ =\ 0\ ,\label{fkfmk}\eea and
 by complex conjugation, \bea \big( \bar\alpha_k\,\bar f_k\ +\
 \bar\alpha_{-k}\,\bar f_{-k}
 \big)\* (p^2-k^2)\QQ\QQ\nn\ =\ \bar\alpha_k\,\bar f_k \*(p-k)\*(p+k)\ +\
 \bar\alpha_{-k}\,\bar f_{-k}\*(p+k)\*(p-k)\ =\ 0\ .
 \label{bfkbfmk}\eea

 These solutions of the $\*$-eigen equation are not real, however.
 The real combinations $\re f_k = (f_k+\bar f_k)/2$ and $\im f_k =
 (f_k-\bar f_k)/(2i)$
 do not satisfy either of
 the $\*$-eigen equations, but do satisfy the \ssee: \bea
 0\ =\ (p^2-k^2)\*\big( f_k\pm\bar f_k \big)\*(p^2-k^2)\QQ\QQ\nn =\
 \big( (p^2-k^2)\*f_k \big)\*(p^2-k^2) \, \pm\, (p^2-k^2)\*
 \big( \bar f_k\*(p^2-k^2)\big)\ .\qq
 \label{fbfkssee}\eea
 This \ssee\ is a fourth-order differential equation,
 whereas the $\*$-eigen equations are second order.  The real
 combinations $(f_k+\bar f_k)/2$ and $(f_k-\bar f_k)/(2i)$ do
 obey a second order equation, of course: \ben
 0\ =\ (p-k)\*\rho\*(p-k)\
 ,\label{iiordk}\een but $(f_{-k}+\bar f_{-k})/2$ and
 $(f_{-k}-\bar f_{-k})/(2i)$ satisfy a different one:\ben
 0\ =\ (p+k)\*\rho\*(p+k)\
 .\label{iiordmk}\een

 So, there is a conflict between reflections and the reality of Wigner
 functions if the free $\*$-eigen equations are to be used. The
 \ssee\ avoids this problem.

\end{document}